\documentstyle[12pt]{article}
\renewcommand \baselinestretch{1.2}

\begin{document}

\def\beq{\begin{equation}}
\def\eeq{\end{equation}}
\def\bce{\begin{center}}
\def\ece{\end{center}}
\def\bea{\begin{eqnarray}}
\def\eea{\end{eqnarray}}
\def\ben{\begin{enumerate}}
\def\een{\end{enumerate}}
\def\ul{\underline}
\def\ni{\noindent}
\def\nn{\nonumber}
\def\bs{\bigskip}
\def\ms{\medskip}
\def\wt{\widetilde}
\def\wh{\widehat}
\def\tr{\mbox{Tr}\, }
\def\brr{\begin{array}}
\def\err{\end{array}}
\def\dsp{\displaystyle}

\hfill U.T.F. 359

\hfill hep-th/9508167

\hfill July, 1995

\vspace*{1cm}

\begin{center}

{\LARGE \bf
On two complementary approaches aiming at the definition of the determinant of
an elliptic partial
differential operator}

\vspace{8mm}

\renewcommand
\baselinestretch{0.5}
\medskip

{\sc E. Elizalde}
\footnote{Permanent address:
Center for Advanced Study CEAB, CSIC, Cam\'{\i} de Santa
B\`arbara, 17300 Blanes,
Spain;
e-mail: eli@zeta.ecm.ub.es} \\
Dipartimento di Fisica, Universit\`a degli Studi di Trento,
I-38050 Povo, Trento,
 Italia \\
Istituto Nazionale di Fisica Nucleare, Gruppo Collegato di Trento

\vspace{15mm}

{\bf Abstract}

\end{center}

We bring together two apparently disconnected lines
 of research (of mathematical and of physical nature, respectively)
  which aim at the definition, through the corresponding zeta function, of
the determinant of  a  differential operator possessing, in general, a complex
spectrum.
 It is shown explicitly how the two lines have in fact converged to a meeting
point at which
the precise mathematical
conditions for the definition of the zeta function and the associated
determinant are
easy to understand from the considerations coming up from the physical
approach, which
proceeds by stepwise generalization starting from the most simple cases of
physical interest.  An explicit formula that establishes the bridge between the
two approaches is
obtained.

\vspace{4mm}


\newpage

\section{Introduction}

It usually happens in different fields of research that the investigations
carried out
by  mathematicians (or by very mathematically minded physicists) on a
particular subject,
and those done by physicists, on the same matter, appear to be quite
disconnected.
 What is even worse, it turns out too often that the results obtained by any of
the two cathegories of researchers remain unknown to the other, even after some
years
of its publication in international journals. It has recently happended to the
author that
several original results on the explicit analytic continuation of a number of
zeta
functions (see \cite{zb1} and the references therein) have been  reobtained
independently by
mathematical colleagues \cite{jl1}.

Simplifying the description a bit, one could distinguish the mathematical from
the
physical approach by saying that the first one proceeds from top to bottom,
aiming always
at the determination of the
most general conditions under which a result or a definition is valid.
Particular
cases and specific applications are usually of secondary importance. The
physical approach,
 on the contrary,
proceeds from below to above, stepwise, at the pace demanded by the neeeds to
solve specific
physical problems of increasing generality and degree of difficulty.

The present investigation has to do with a  clear example of this issue, which
corresponds to the
important problem of the  the definition of the concept of determinant of an
elliptic
partial differential operator (PDO) exhibiting (in general) a complex spectrum.
Two different approaches to this problem exist in the literature, possessing
respectively the
characteristics that we have just described above.  Not surprisingly, both
lines of research
 ---one of mathematical nature, coming from above, and the other pursued by
physicists, coming from
below---  will be shown here to converge explicity
 to a meeting point at which
the precise mathematical
conditions for the definition of the concept of determinant will be
easy to understand from the considerations coming from the physical approach,
which
proceeds by rather straightforward generalization starting from the most simple
cases of
physical interest.  A formula that establishes the bridge between the two
approaches will
be obtained.

 In fact, the definition of the concept of determinant of a general elliptic
PDO
is still an open problem, in the sense that there are several possible
definitions of this
concept and that none of them has been proven to extend to the whole class of
such
operators. In fact, for invertible operators of the form $A=I+K$, with $K$ an
operator of
trace-class acting on a Hilbert space (of infinite dimension, in general), the
most
popular definition of determinant is the one due to Fredholm
\begin{equation}
{\det}_{Fr}(I+K)= I+ \sum_{n=1}^\infty \tr\left(\wedge^n K\right)
\label{t3.1}
\end{equation}
This series is absolutly convergent for the operators considered, and the
Fredholm
determinant has the usual properties of the determinant for finite-dimensional
matrices, in
particular the  fundamental property:
\begin{equation}
\det (AB) = \det (A) \, \det (B).
\label{t3.2}
\end{equation}
The Fredholm determinant reduces to the ordinary definition when the dimension
of the Hilbert space
is finite.

 But maybe the most useful definition of determinant for elliptic PDOs is the
one proposed by Ray
and Singer in the early seventies \cite{rs123}, which makes use of the concept
of zeta function
of an operator, $\zeta_A (s)$ (a generalization, on its turn, of the Riemann
zeta function,
$\zeta_R (s)= \sum_{n=1}^\infty n^{-s}$, Re $s >1$).
The zeta (or Ray-Singer) determinant is defined as
\begin{equation}
{\det}_\zeta (A) = \exp \left( - \left. \frac{d}{ds} \zeta_A(s)\right|_{s=0}
\right),
\label{t3.3}
\end{equation}
and the (formal) definition of $\zeta_A(s)$ is the following
\begin{equation}
\zeta_A (s) =  \sum_{n}  \lambda_n^{-s},
\label{t3.4}
\end{equation}
being $\lambda_n$ the eigenvalues of $A$ and $s$ a complex variable. It is
clear that there are
many problems associated with this definition (for instance, in general the
spectrum is unknown,
and there can be zero eigenvalues). Being more precise, for a non-negative,
selfadjoint operator
$A$ one proceeds by  analytically continuing  in the complex $s$-plane ---from
a region Re $s >
s_0$, with $s_0$ some abscissa of convergence, to the region Re $s \leq s_0$---
the function given
by the series  (\ref{t3.4}) above with the zero eigenvalues excluded from the
sum. And this can be
carried out explicitly in many different situations, increasingly complicated
 (see \cite{zb1} and references therein).

Now we shall summarize the mathematical approach to the question.
 Presently, the most general conditions for
the definition of the zeta function of an elliptic PDO are known to be the
following: (i) the order
$p$ of $A$ must be real and different from zero; (ii) there must exist a
conical neighborhood $U$
of a ray $L$ from the origin in the spectral plane, such that the principal
symbol $a_p(x,\xi)$ of
the operator $A$ has no eigenvalue in $U$ (for all points $x$ of the manifold
$M$, where $A$ acts,
 and for any
$\xi \in T_x^*M$,  with $\xi \neq 0$).  Of course in the finite-dimensional
case this condition is
obviously satisfied. Also, it is immediate that by  multiplying the operator
$A$ with a convenient
 constant
$c \in $ {\bf C}, $|c|=1$, the ray $L$ can be taken to be (without restriction)
the negative real
axis {\bf R}$_-$ (by working then with the operator $A_1 =cA$, instead of $A$).

However, it turns out that even if $\zeta_A(s)$, $\zeta_B(s)$ and
$\zeta_{AB}(s)$ exist, for the
elliptic differential operators $A$ and $B$, in general
\begin{equation}
{\det}_\zeta (AB) \neq {\det}_\zeta (A) \, {\det}_\zeta (B).
\label{t3.5}
\end{equation}
This has led Friedlander \cite{fried1} and Konsevich and Vishik \cite{kv1} to
consider the
multiplicative anomaly
\begin{equation}
F(A,B) \equiv \frac{{\det} (AB) }{ {\det} (A) \, {\det} (B)}.
\label{t3.6}
\end{equation}
For the case when $A$ and $B$ are positive-definite elliptic PDOs of positive
order, it has been
shown in \cite{fried1} that the anomaly is equal to one and that the three zeta
functions can be
defined with the help of a cut in the spectral plane on  {\bf R}$_-$. In
\cite{kv1}, on the
other hand, an expression of $F(A,B)$ in terms of the symbols of the operators
has been obtained
and it has been proven that $F=1$ for a certain class of PDOs in
odd-dimensional manifolds which
generalizes the class of elliptic PDOs. Within this class, a new definition of
determinant has been
introduced, which is valid even for zero-order operators, thus extending
non-trivially the
definition of zeta determinant. In particular, the definition is valid for
invertible
pseudo-differential operators close to positive self-adjoint ones, getting then
back to the
definition of zeta determinant. This much for the mathematical approach.

As starting point for the more physical approaches to the definition of a
determinant let us take
the papers of Schwarz \cite{schw2}, initiating in the late seventies
\cite{schw1},
 in which use is made of the
concept of Ray-Singer  (zeta) determinant for the calculation of the partition
function of a gauge
field theory. To fix up ideas, in  topological field theory, it is given by the
expression
\begin{equation}
Z(\beta )= \int_\Gamma {\cal D} \omega \, e^{-\beta S(\omega)},
\label{t3.7}
\end{equation}
where $\omega$ are fields on a manifold $M$, $\Gamma$ is some linear space of
fields  and $S$ is
the topological action functional.  The constant $\beta$ was considered in
\cite{schw1} to be real
($\beta =1$). Later, Witten \cite{witt1} had to calculate the case when $\beta$
is purely
imaginary and, very recently, Adams and Sen \cite{as1} have attacked the
situation
 of arbitrary complex
 $\beta$. Those are in brief the steps of the development in physics. For
ulterior uses,
it will be enough to consider the case of a compact manifold, $M$ without
boundaries and oriented,
and a quadratic action $S$, e.g. $S(\omega) = <\omega, A\omega>$, with respect
to some Euclidean
metric on $M$.

In this case,  the definition of the zeta function of $A$, Eq. (\ref{t3.3}),
 written under the form of a  Mellin transform, is
\begin{equation}
\zeta_A (s) =  {\sum_{n=1}^\infty }' \lambda_n^{-s}= \frac{1}{\Gamma}
\int_0^\infty dt \, t^{s-1}
\tr \left( e^{-tA} - \Pi_{Ker}\, A \right).
\label{t3.8}
\end{equation}
Moreover, for $B$ an operator of order zero (multiplication by a function) one
has the
asymptotic expansion
\begin{equation}
\tr \left( B\,  e^{-tA}  \right) \sim \sum_k c_k(B|A)\, t^k, \qquad
t\rightarrow 0^+,
\label{t3.9}
\end{equation}
where the $c_k(B|A)$ are the heat-kernel coefficients, first calculated by
Seeley
\cite{seel1} and subsequently by
a long list of authors, in different situations (for very new results and a
list of references,
see \cite{bek1}).
It can be shown, using these two equations, that the zeta-function of a general
non-negative
elliptic operator (self-adjoint and of positive order) is meromorphic in the
complex $s$-plane,
and analytic at $s=0$.  For $\beta =1$ in Eq. (\ref{t3.7}), it was proven in
Ref. \cite{schw1} ---by
extending the Faddeev-Popov trick for zero modes--- that
\begin{equation}
Z(\beta =1 )= ({\det}_\zeta \wt{A})^{-1/2} \prod_{j=1}^N \left(\det ( T_j^+
T_j)\right)^{(-1)^{j+1}/2},
\label{t3.10}
\end{equation}
being  $\wt{A} \equiv A - \Pi_{Ker} \, A$. Here $\{ T_j \}$ is a sequence of
linear operators
which constitute a resolvent of $S$ ($T:\Gamma_j \rightarrow \Gamma_{j-1}$,
linear spaces, with
$S(f+T_1g)=S(f)$, $T_{j-1} T_j =0$, $j=0,1,\ldots, N$). Notice that when
generalizing
 Eq. (\ref{t3.10}) to $\beta \neq 1$,  all the
dependendence on $\beta$ will appear in the first of the determinants on the
rhs, and thus we
 are not going to consider further the other factors, corresponding to the
resolvent,
 in the analysis to follow.

The case $\beta = -i$ was calculated by Witten in 1989 for the case of the
Chern-Simons theory
 \cite{witt1}. Taking as starting point the
previous results, he just had to obtain the additional phase of the
determinant. He proceded
by directly computing the integral in an orthonormal basis of eigenfunctions
for $A$, $x_i$, i.e.
\begin{equation}
\prod_n \int_{-\infty}^\infty \frac{dx_n}{\sqrt{\pi}} \, e^{i\lambda_nx_n^2}.
\label{t3.11}
\end{equation}
Using an $\epsilon$-regularization, one obtains, for each of the factors
\begin{equation}
 \int_{-\infty}^\infty \frac{dx}{\sqrt{\pi}} \, e^{i\lambda x^2} =
\lim_{\epsilon \rightarrow 0}
\int_{-\infty}^\infty \frac{dx}{\sqrt{\pi}} \, e^{i\lambda x^2} \,
e^{-\epsilon x^2} =
\frac{1}{|\sqrt{\lambda}|} \exp \left( \frac{i\pi}{4} \mbox{sgn} \, \lambda
\right),
\label{t3.12}
\end{equation}
and for the whole determinant
\begin{equation}
\left({\det}_\zeta ( -i \wt{A})\right)^{-1/2} =\left({\det}_\zeta (
\wt{A})\right)^{-1/2} \, \exp
 \left( \frac{i\pi}{4} \eta_A (0) \right),
\label{t3.13}
\end{equation}
where
\begin{equation}
\eta_A (s) = \sum_n  \mbox{sgn} (\lambda_n) \, |\lambda_n|^{-s}
\label{t3.14}
\end{equation}
is the eta function of $A$ (also called Atiyah-Patodi-Singer's eta invariant
\cite{ati1} in the
limit $s\rightarrow 0$ and affected by a factor $1/2$).

The more general case of $\beta$ arbitrary complex has been investigated in
Ref. \cite{as1}.
 By extending the
considerations of Witten, these authors obtain the formula
\begin{equation}
\left( \det(\beta\widetilde{A})\right)^{-1/2}
=\exp \left\{-\frac{i\pi}{4}\left[\left(\frac{2\theta}{\pi}\mp1\right)\zeta\;
\pm\;\eta\right] \right\} \,|\beta|^{-\zeta/2}\, \left( \det
|\widetilde{A}|\right)^{-1/2},
\label{t3.15}
\end{equation}
 where $\zeta \equiv \zeta_{|A|} (0)$, $\eta \equiv \eta_{A} (0)$, and where
the $\pm$ signs
correspond, respectively, to the two situations of $\beta$ belonging to the
upper or lower half of
the complex plane, respectively, i.e.
\begin{equation}
\beta = |\beta| \, e^{i\theta} \quad  \left\{ \brr{ll} +: & 0\leq \theta \leq
\pi, \\
  -: & \pi\leq \theta \leq 2\pi. \err \right.
\label{t3.16}
\end{equation}
As the authors themselves recognize, there is an ambiguity in this expression
for
$\beta$ real, and thus the formula only generalizes Witten's case, but
{\it not} Schwarz's one! As anticipated before, the problem is a clear example
of
 the lack of connection between physicists and mathematicians. The authors of
\cite{as1} seem to
be in fact unaware of the strong results that have been summarized above
\cite{kv1}.

We shall now bring together the above two lines of research, the mathematical
one described
 in the first
paragraphs ---that tries to extend the definition of the zeta function and of
the associated
determinant to the most general class of elliptic PDOs possible--- and the
physical one, that
approaches this goal from below, step by step,
starting from the simple cases (\ref{t3.10}) and (\ref{t3.13}),
 as the situations  in actual physical
theories demand (e.g., nowadays, topological field theories).  It is certainly
clear that the
powerful results
of the first line of investigation contradict the statement that the zeta
function has not
been defined in the literature for operators with a negative spectrum
\cite{as1}. This
assertion might only become true if one would
 add {\it in the `physical' literature}. We have seen that a much
more general definition of zeta function have been given for operators with a
complex spectrum,
 under the
 very specific condition of the existence of a conical neighborhood of a
certain ray from the
origin,
in the complex spectral plane, where no eigenvalue is present.

It is quite remarkable to see how the condition of existence of the conical
neighborhood
 over a ray in
the spectral plane arises in an absolutely natural way from the physical,
stepwise approach to the
problem. This is obtained by simply complementing the techniques of Schwarz and
Witten with a
 well-defined analytic
continuation on the complex plane of the power function of the constant
$\beta$.
  Oddly enough, the final result and, in
particular, the definitions themselves of the zeta function and of the
associated determinant will
turn out to be independent of the position of this ray in the spectral plane.

In fact, the crucial point in the whole procedure is a quite simple one, to be
learned in any
standard course on complex variable. It consists namely on  the proper
definition of the analytic
 continuation of the
power function, $z^a$ for $a\in$ {\bf R}, from $z \in$ {\bf R}$_+$ to complex
values of $z$, i.e.,
\begin{equation}
z = |z| \, e^{i\theta} \  \ \longrightarrow  \ \  z^a = |z|^a \, e^{ia\theta},
\qquad  a\in
\mbox{\bf R}.
\label{t3.17}
\end{equation}
The key point here is the appearance of a cut from the origin in the complex
$z$-plane, that
can be fixed at will by selecting  the interval of variation of the argument,
that is,
\begin{equation}
\theta \in [-\gamma, 2\pi -\gamma], \qquad  0 <\gamma < 2\pi.
\label{t3.18}
\end{equation}
Thus,  the cut (in other words, the ray $L$ from the origin) has been here
chosen to be
\begin{equation}
L= \{ z \in \mbox{\bf C}\, | \, Arg (z)=\gamma\}.
\label{t3.19}
\end{equation}
 Summing up, the angle $\gamma$ defines the ray $L$ from the origin on the
complex plane
which contains the points that are not reached from the specific analytic
continuation chosen, in
the sense that $z^a$ is not (uniquely) defined for those points $z$ (a double
phase appears). Of
course, in the books one will find that this ray can be always taken to be $L=$
{\bf R}$_-$, and
that  this is the most `natural' choice. Also, as we have seen before, by
simple multiplication by a
complex constant of modulus one, one can reduce the whole discussion for a
general operator to this
standard case. But we, physicists, should be extremely careful with this kind
of general
mathematical considerations. For a given
operator $A$ the ray {\it cannot} always be put on the negative real axis. A
different thing is
that the general discussion concerning the operator $A$, with a cut suitably
fixed at
$L= \{ z \in \mbox{\bf C}\, | \, Arg (z)=\gamma\}$, can in fact be
reduced to an {\it equivalent} discussion of the operator $A_1=A \, e^{i(\pi
-\gamma)}$ which has the
`standard' cut  at  $L=$ {\bf R}$_-$.

And this is the reason why the formula (\ref{t3.15}) does not work for the most
simple case, when
the spectrum of $A$
has a real negative part. Let us now put everything together and derive the
desired
formula. Recall that now  $A$ is an operator with a general real spectrum.
 Corresponding to the partition of the identity operator into the different
parts of the spectrum,
$I=\Pi_{Ker} + \Pi_+ +\Pi_-$, let us write $\wt{A} = A_+ +A_-$ (remember that
$\wt{A}
=(I-\Pi_{Ker}) A$). $\beta$ will be a general complex number and we can proceed
with the
straightforward calculation \cite{as1} (we shall from now on drop out the
subscript $\zeta$ from the
det)
\bea
\left(\det (\beta \wt{A})\right)^{-1/2} &=& \left(\det (\beta
A_+)\right)^{-1/2}
 \left(\det (\beta  A_-)\right)^{-1/2} = \left(\det (\beta  A_+)\right)^{-1/2}
 \left(\det [(-\beta )(- A_-)]\right)^{-1/2} \nn \\
&=& (-1)^{\dsp -\frac{1}{2} \zeta_{-A_-} (0)} \beta^{\dsp -\frac{1}{2}
[\zeta_{A_+} (0) +
 \zeta_{-A_-} (0)]} \left(\det |\wt{A}|\right)^{-1/2}.
\label{t3.20}
\eea
By using  the identities
\beq
\zeta_{|A|} (s) = \zeta_{A_+} (s) +  \zeta_{-A_-} (s), \qquad
 \eta_{A} (s) = \zeta_{A_+} (s) -  \zeta_{-A_-} (s),
\label{t3.21}
\eeq
we get the relations
\beq
\zeta_{A_+} (0) =\frac{1}{2} \left( \zeta + \eta \right), \qquad
 \zeta_{-A_-} (0) =\frac{1}{2} \left( \zeta - \eta \right),
\label{t3.22}
\eeq
wherefrom we obtain the final formula
\beq
\left(\det (\beta \wt{A})\right)^{-1/2}= (-1)^{ -( \zeta - \eta ) /4}
\beta^{-\zeta /2}  \left(\det |\wt{A}|\right)^{-1/2}.
\label{t3.23}
\eeq
In this formula the powers of $(-1)$ and $\beta$ are to be taken in the
$\gamma$-analytic
 continuation
chosen for the power function, which is the one appropriate in order that all
the powers are
uniquely defined. In the cases considered in topological field theory, one is
free to choose any
value for $\gamma$ that is not a multiple of $\pi /2$. In a more general case,
 with $\beta A$ general
elliptic, one must guarantee in advance that this choice exists, in order to be
able to
 define the zeta
function.  This is the content of the powerful mathematical theorem, as seen
from the physical
viewpoint. Another important point for the formula
(\ref{t3.23}) to be valid is that $\zeta_{|A|} (s)$ and
$ \eta_{A} (s)$ must be analytic at $s=0$. This is certainly true in the
ordinary conditions of a
non-negative elliptic operator (see the paragraph before Eq. (\ref{t3.10})).
That such is also the
case for the situations of interest in topological field theories (forms of
degree $m$ in a manifold
$M$ of odd dimension $2m+1$) has been beautifully shown in \cite{as1}. Again,
this is the case
for which in the mathematical theory the determinant can be defined:
for an automorphism of a vector bundle on an odd-dimensional manifold
acting on global sections of this vector bundle.
Also, a natural trace has been introduced in \cite{kv1} for PDOs of odd class
on an odd-dimensional closed $M$.

To finish, let us consider some particular uses of the formula (\ref{t3.22}),
that connect with
results previously obtained in the literature. The most simple case is  when
$\beta$ is real and
positive.
Then only the power of $(-1)$ needs analytical continuation in Eq.
(\ref{t3.23}).
On the other hand, for  $\beta$ real and negative the formula reads
\beq
\left(\det (\beta \wt{A})\right)^{-1/2}= (-1)^{( \zeta + \eta) /4}
(-\beta)^{-\zeta /2}  \left(\det |\wt{A}|\right)^{-1/2}
\label{t3.24}
\eeq
where,  again,  only the power of $(-1)$ needs analytical continuation.
 For $\beta$ purely imaginary we
reobtain Witten's result
\bea
\left(\det (\beta \wt{A})\right)^{-1/2} & = & (-1)^{- ( \zeta - \eta) /4 +
\zeta /2} \,
 i^{-\zeta /2} \,  |\beta|^{-\zeta /2}\,  \left(\det |\wt{A}|\right)^{-1/2} \nn
\\
 &= & e^{i\pi \eta /4} \, |\beta|^{-\zeta /2} \, \left(\det
|\wt{A}|\right)^{-1/2}. \label{t3.25}
\eea
It is nice to observe that this result is here just a particular case of the
ordinary zeta
function procedure for defining the determinant, nothing extra has to be
introduced, but the
overall consideration of performing analytic continuations properly. Eq.
(\ref{t3.25})
 was obtained in \cite{witt1} using an $\epsilon$ regularization factor and for
$|\beta|=1$ (it
was already noticed there that any reasonable regularization should yield the
same result).
Finally, for $L=$ {\bf R}$_-$ we recover the formula (\ref{t3.15}) of Adams and
Sen
 \cite{as1}, which is certainly
unambiguous as a generalization of the Witten case ($\beta$ pure imaginary)
but not of the $\beta =1$  case of Schwarz to $\beta \in$ {\bf R}$_-$.

Summing up, formula (\ref{t3.23})
encompasses all those situations, extends them to $\beta$ arbitrary complex,
and establishes
 an explicit connection with the general definition of zeta-function
determinant and with the
more far reaching considerations in \cite{kv1}.

 \vspace{5mm}


\noindent{\large \bf Acknowledgments}

It is a pleasure to thank the members of the Department of Theoretical Physics
of the
 University of Trento for
warm hospitality, specially Sergio Zerbini, Luciano Vanzo, Guido Cognola,
Ruggero Ferrari
 and Marco Toller.
This work has been supported by DGICYT (Spain), by CIRIT (Generalitat de
Catalunya) and by INFN
(Italy).





\end{document}